# Post-Mortem Human Iris Segmentation Analysis with Deep Learning

Afzal Hossain[1], Tipu Sultan[2], Stephanie Schuckers[3]

## Abstract

*Iris recognition is widely used in several fields such as mobile phones, financial transactions, identification cards, airport security, international border control, voter registration for living persons. However, the possibility of identifying deceased individuals based on their iris patterns has emerged recently as a supplementary or alternative method valuable in forensic analysis. Simultaneously, it poses numerous new technological challenges and one of the most challenging among them is the image segmentation stage as conventional iris recognition approaches have struggled to reliably execute it. This paper presents and compares Deep Learning (DL) models designed for segmenting iris images collected from the deceased subjects, by training SegNet and DeepLabV3+ semantic segmentation methods where using VGG19, ResNet18, ResNet50, MobileNetv2, Xception, or InceptionResNetv2 as backbones. In this study, our experiments demonstrate that our proposed method effectively learns and identifies specific deformations inherent in post-mortem samples and providing a significant improvement in accuracy. By employing our novel method MobileNetv2 as the backbone of DeepLabV3+ and replacing the final layer with a hybrid loss function combining Boundary and Dice loss, we achieve Mean Intersection over Union of 95.54% on the Warsaw-BioBase-PostMortem-Iris-v1 dataset. To the best of our knowledge, this study provides the most extensive evaluation of DL models for post-mortem iris segmentation.*

## 1. Introduction

Iris recognition, known for its unique texture patterns, provides precise and secure personal authentication [1], [2]. It has been a solid biometric identification means for government ID cards, banking transactions, border crossing, mobile application, FBI's Next Generation Identification system [3], [4], [5]. Due to the COVID-19 pandemic, iris recognition has become more reliable biometric recognition method as it is contactless and hygienic. Recently post-mortem iris recognition has also gained attention from the biometric community because of the importance in forensics, investigations involving criminal activities or military operations on the battlefield. As a biometric identifier, iris has high temporal stability compared to other biometric modes [6]. Trokielewicz et al. [7] suggest that conventional algorithms can provide accurate matches for iris samples acquired up to 17 days after death, particularly when bodies are maintained in mortuary conditions. Similarly, Sauerwein et al. [8] demonstrated in their experiments that irises remain viable for up to 34 days post-mortem, particularly when cadavers are exposed to outdoor conditions, such as during the winter. There are some other research groups who have been investigating this area and have been proposing Presentation Attack Detection (PAD) techniques targeted at identifying cadaver iris presentations to the sensor [9], [10]. Post-mortem iris recognition is also being considered as a potentially valuable method for forensic procedures as it is reliable and fast. The forensic community has recognized several scenarios where the speed of post-mortem iris recognition proves valuable, appreciating its usefulness. One such scenario involves the more accurate matching of ante-mortem samples with post-mortem data obtained at crime scenes and mass fatality incidents. In case of any accident, rapid registration of the body at the scene is necessary to facilitate subsequent tracking and proper dispatch to either the family or a mortuary. In case of mass fatality accidents, where time is critical, forensic practitioners in several countries have chosen to utilize iris recognition over the slower DNA identification method.

It has been found that while the iris recognition algorithms work good for the living persons, the performance of the algorithms deteriorate when confronted images from deceased subjects [11], [12]. The effectiveness of iris recognition detrimentally impacted as the deterioration continue over time since death elapses, due to significant distortions of the iris and the cornea caused by post-mortem decay processes. Inconsistent image segmentation is frequently cited as a common factor contributing to the deterioration of iris recognition algorithm performance, particularly when dealing with challenging samples, like post-mortem samples. Post-mortem decay occurring at the cellular level gradually results in macroscopic changes within the eye. These changes include deviations from the circularity of the pupil,

1. Afzal Hossain; PhD Student, Electrical and Computer Engineering, Clarkson University, 8 Clarkson Avenue, Potsdam, NY13676, US, afhossa@clarkson.edu
2. Tipu Sultan; PhD Student, Mechanical and Aerospace Engineering, Clarkson University, 8 Clarkson Avenue, Potsdam, NY13676, US, sultant@clarkson.edu
3. Stephanie Schuckers; Professor, Electrical and Computer Engineering, Clarkson University, 8 Clarkson Avenue, Potsdam, NY13676, US, sschucke@clarkson.edu1

wrinkles on the cornea leading to additional specular reflections, and alterations in the texture of the iris [7]. In this situation, the accurate execution of the segmentation stage is essential to ensuring the high accuracy of iris recognition. This accuracy relies on encoding the actual texture of the iris, rather than the surrounding portions of the eye to mitigate the death effects on iris. Therefore, the accuracy and robustness of iris segmentation is very crucial as it can directly impact subsequent phases such as iris extraction, verification, and recognition [13].

Ensuring the accuracy and precision of iris segmentation has posed a considerable challenge in uncontrolled environments for many years. Several authors have put forward a range of distinct approaches and strategies to tackle and overcome these complex challenges. Although traditional iris segmentation methods have made significant contributions, recent advancements in DL techniques have demonstrated notable enhancements in iris segmentation performance, particularly for post-mortem iris images. To the best of our knowledge, no other paper or published research has conducted a comprehensive evaluation of DL models for post-mortem iris segmentation. In this paper we conduct experiments integrating pretrained models like VGG19, ResNet18, ResNet50, MobileNetV2, Xception, and InceptionResNetV2 with DeepLabV3+ and SegNet. We assess various accuracy metrics, with particular emphasis on Mean Intersection over Union, which is regarded as the most suitable metric for iris segmentation evaluation. Additionally, we customize the loss function of pretrained models by substituting it with the Dice loss function. This choice is made because the Dice loss is closely associated with mean IoU and frequently employed in analogous scenarios for image segmentation tasks. In our study, we perform further experiments with some other loss functions. Experimental results indicate that our methods show significant improvements compared to post-mortem iris segmentation outcomes obtained from other methods.

## 2. Related Works

Iris segmentation and recognition for living individuals has been well-established for a long time. However, the recognition of deceased individuals using their iris patterns has traditionally been considered impossible. Even prior to significant experimental studies on post-mortem iris recognition, the researchers doubted its feasibility [10]. The previous assumption suggests that soon after death, the pupil dilates considerably and the cornea becomes cloudy which has influenced subsequent researchers [5]. Consequently, conclusions have been drawn, such as the iris decaying only a few minutes after death [14]. Sansola provided initial evidence demonstrating the feasibility of matching perimortem (acquired just before death) and postmortem irises [11]. The study involved photographing irises of 43 deceased subjects at various post-mortem intervals using the IriShield M2120U iris recognition camera. They utilized IriCore matching software and observed correct matching results for at least 70% of cases when only postmortem irises were compared, with the success rate varying based on the time elapsed after death. Their experimental results showed false non-match rates ranging from 19% to 30%, with no instances of false matches, contingent on the time elapsed since death.

Boyd et al. [15] conducted a comprehensive survey summarizing various aspects of experimenting with post-mortem iris recognition. Saripalle conducted a study on ex-vivo eyes obtained from domestic pigs, examining their biometric capacity as tissue degradation progressed following removal from the cadaver [16]. Their research revealed that the irises lose their biometric capabilities within 6 to 8 hours after death, emphasizing the swift degradation of ex-vivo eyes. Bolme et al. [17] documented that post-mortem irises can be used for segmentation and recognition are viable only for a short time after death, especially when bodies are exposed outdoors during the summer. This is attributed to the intense and rapid decomposition processes occurring within the eye. However, Trokielewicz et al. [7] demonstrated that correct segmentations and matches can still be expected even after 17 days and their extended study [18], as well as Sauerwein et al. [8] propose that accurate matches may still be achievable even three to five weeks after death, particularly in mortuary and winter-time outdoor conditions. Trokielewicz et al. [12] introduced the first publicly available dataset of near-infrared and visible-light post-mortem iris images where their experimental results show that under temperatures below 42°F, the iris can still successfully serve as a biometric identifier for 27 hours after death.

Several authors in the aforementioned works have employed DL based methods, which have recently demonstrated significant potential in addressing specific computer vision tasks, such as natural image classification and dense labeling image segmentation [19], [20] . Numerous attempts have been made regarding the applications of iris segmentation employing neural networks, primarily focusing on enhancing the segmentation of challenging and noisy iris images. Broussard and Ives utilized neural networks to train a Multi-Layer Perceptron (MLP) for discerning the most discriminative measurements and iris regions by identifying and labeling unwrapped polar iris image pixels as either iris or non-iris [21]. Marra et al. [22] employed Convolutional Neural Networks (CNNs) to classify iris images acquired from various imaging devices. Liu et al. [23] investigated the use of Hierarchical Convolutional Neural Networks (HCNNs) and Multi-Scale Fully Convolutional Neural Networks (MFCNs) to enhance the segmentation of noisy iris images. Trokielewicz et al. [24] proposed a Deep Convolutional Neural Network (DCNN) based post-

1. Afzal Hossain; PhD Student, Electrical and Computer Engineering, Clarkson University, 8 Clarkson Avenue, Potsdam, NY13676, US, afhossa@clarkson.edu
2. Tipu Sultan; PhD Student, Mechanical and Aerospace Engineering, Clarkson University, 8 Clarkson Avenue, Potsdam, NY13676, US, sultant@clarkson.edu
3. Stephanie Schuckers; Professor, Electrical and Computer Engineering, Clarkson University, 8 Clarkson Avenue, Potsdam, NY13676, US, sschucke@clarkson.edu2

mortem iris segmentation model which was created on the SegNet architecture comprising a classification layer, a corresponding decoder, and an encoder. The authors train the network with ten subject-disjoint train/test data splits, evaluating segmentation accuracy using Intersection over Union (IoU). Compared to the conventional OSIRIS method, the DCNN-based solution achieved higher segmentation accuracy, with an average IoU of 88.53% compared to OSIRIS's 73.58%. This represents an average improvement of 12.8% over the conventional approach, with consistent outperformance across all data splits.

In our paper, we conduct experiments to investigate the effectiveness of DL models for post-mortem iris segmentation, mainly focusing on improving the lacking of these models. We integrate pretrained models like VGG19, ResNet18, ResNet50, MobileNetV2, Xception, or InceptionResNetV2 with the semantic segmentation models DeepLabV3+ and SegNet. Additionally, we modify the loss function layer of the pretrained models and analyze the results. Our experimental findings reveal methods that can improves post-mortem iris segmentation accuracy with reduced number of parameters and computational complexity. The remainder of the paper is organized as follows:

In Section 3, we provide an explanation of the semantic segmentation models, pretrained models, and loss functions employed in our methods. Section 4 details the iris image databases used and the corresponding ground-truth masks, along with accuracy metrics, training, and evaluation procedures with experimental results and discussions. Finally, Section 5 summarizes our study.

## 3. Network Architecture

Using pretrained deep neural networks as a backbone with semantic segmentation networks, we blend advanced feature extraction with precise segmentation. This combination creates a strong and accurate system for iris segmentation. In our approach, we employ SegNet and DeepLabV3+ (provided by MATLAB Neural Network Toolbox), two distinct models for semantic segmentation. We substitute the usual encoder of the SegNet model with a pre-trained VGG19 network. Similarly, for DeepLabv3+, we use ResNet18, ResNet50, MobileNetv2, Xception, or InceptionResNetv2 as backbones. We acknowledge the Mathworks Deep Learning Toolbox Team for making all the pretrained models publicly available that we use in our experimental study. In addition, we conduct experiments replacing the original loss function layer with several other loss functions. This section contains a description of the networks and Dice loss function that we employ in our experiments.

### 3.1 Semantic Segmentation Methods:

*3.1.1 SegNet: A Deep Convolutional Encoder-Decoder Architecture for Image Segmentation [25]:* SegNet is a deep neural network architecture designed for semantic segmentation tasks. SegNet is made up of a 13 convolutional layers encoder network, with a corresponding decoder layer for each encoder layer. For translation invariance, the encoder network uses max-pooling and sub-sampling. However, it only retains max-pooling indices to save memory. With the help of these indices, the decoder produces dense feature maps by upsampling the input feature maps and employing trainable decoder filter banks. A softmax classifier individually calculates class probabilities for each pixel.

*3.1.2 Encoder-Decoder with Atrous Separable Convolution for Semantic Image Segmentation [26]:* Google invented DeepLabv3+, an upgrade to DeepLab3 *[27]* where the first method of this DeepLab series is Chen et al. *[28]*. Atrous convolution and depth wise separable convolution are two essential convolution techniques that DeepLabv3+ presents. These strategies are crucial to reduce computational complexity in deep convolutional neural networks and capture multi-scale information. The use of DeepLabv3+ as an encoder module place special emphasis on the capacity to extract features at an arbitrary resolution. The authors suggest a decoder module that improves object segmentation information. They also make changes to the Xception model, creating the "Modified Aligned Xception," which is specifically designed for semantic image segmentation. These changes consist of a deeper Xception network with effective computation and memory usage, depthwise separable convolution in place of max pooling, extra batch normalization, and ReLU activation for improved performance. An encoder-decoder architecture of DeepLabv3+ is shown in Figure 1.

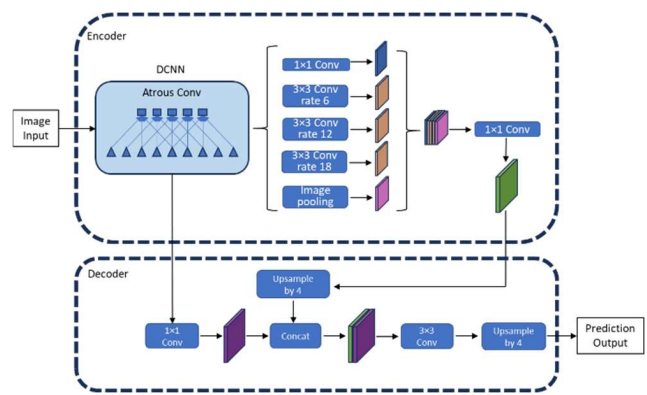

**Figure 1.** Encoder-decoder architecture of DeepLabv3+.

1. Afzal Hossain; PhD Student, Electrical and Computer Engineering, Clarkson University, 8 Clarkson Avenue, Potsdam, NY13676, US, afhossa@clarkson.edu
2. Tipu Sultan; PhD Student, Mechanical and Aerospace Engineering, Clarkson University, 8 Clarkson Avenue, Potsdam, NY13676, US, sultant@clarkson.edu
3. Stephanie Schuckers; Professor, Electrical and Computer Engineering, Clarkson University, 8 Clarkson Avenue, Potsdam, NY13676, US, sschucke@clarkson.edu3

## 3.2 Pretrained Deep Neural Networks (Backbones):

In this study, we explore various neural network architectures and their implications for large-scale image recognition, incorporating advanced models like VGG-16, ResNet-18, ResNet-50, MobileNetV2, Xception, and Inception-ResNet-v2, each serving as the backbone for either SegNet or DeepLabv3+ segmentation frameworks. The VGG-19 model [29], with its 19-layer depth and exclusive use of 3×3 convolutional layers, eschews Local Response Normalization in favor of max-pooling and ReLU activations, culminating in a softmax layer for classification. This setup, when integrated with SegNet, is designated as model "$M_1$", tailored for high fidelity in segmenting intricate scenes from standardized 224×224 RGB inputs. Exploring further, the DeepLabv3+ framework employs deep residual learning principles with ResNet backbones, where ResNet-18 and ResNet-50 models [30] ("$M_2$" and "$M_3$", respectively) aim to learn residual functions to mitigate the degradation problem inherent in deep networks, leveraging identity mapping via shortcut connections for improved convergence. These models, pretrained on the expansive ImageNet dataset, have developed rich feature representations, enabling precise classification across a wide array of object categories. The MobileNetV2 architecture [31], designated as model "$M_4$" within the DeepLabv3+ framework, introduces a approach with its 53-layer depth, utilizing bottleneck blocks and depth-separable convolutions to balance performance and computational efficiency through adjustable hyperparameters like image resolution and width multiplier. Model "$M_5$" emerges from the integration of the Xception architecture [32] with DeepLabv3+, where the Xception model, stretching across 71 layers, champions depthwise separable convolutions within a linear stack of residually connected modules, advocating for the decoupling of cross-channel and spatial correlations in convolutional feature maps. Lastly, the Inception-ResNet-v2 architecture [33], forming the backbone of model "$M_6$" within the DeepLabv3+ framework, amalgamates the best of Inception blocks with residual connections, aiming for a harmonious blend of computational efficiency and model simplicity without sacrificing depth or performance, all the while trained on a vast corpus of ImageNet data to ensure a broad and robust classification capability. This eclectic mix of models, each with its distinct architectural philosophy and integration with segmentation frameworks, underscores the diverse strategies in pushing the boundaries of image segmentation accuracy.

## 3.3 Loss functions:

**Dice Loss:** The Dice loss function, based on the Sørensen-Dice similarity coefficient, measures the overlap between two segmented images. The generalized Dice loss function L, for the loss between one image Y and the corresponding ground truth T is expressed as:

$$L = 1 - \frac{2 \sum_{n=1}^{N} w_n \sum_{m=1}^{M} Y_{nm} T_{nm}}{\sum_{n=1}^{N} w_n \sum_{m=1}^{M} Y^2_{nm} + T^2_{nm}} \quad (1)$$

N represents the number of classes, M denotes the number of elements along the first two dimensions of Y, and $w_n$ is a class-specific weighting factor controlling each class's contribution to the loss. These weightings are crucial for mitigating the influence of larger regions on the Dice score, facilitating the network's learning process for segmenting smaller regions effectively. Typically, $w_n$ corresponds to the inverse area of the expected region:

$$w_n = \frac{1}{(\sum_{m=1}^{M} T_{nm})^2} \quad (2)$$

This loss function employs a variant of the generalized Dice Loss function [34], [35], incorporating squared terms to ensure a derivative of 0 when the prediction aligns with the ground truth, thereby stabilizing training and promoting accurate segmentation convergence [36].

## 4. Results and Analysis

### 4.1 Datasets:

In our experiment, we have employed three distinct datasets. For training our model, we utilized the Warsaw-BioBase-PostMortem-Iris-v1 dataset and the corresponding ground truth masks publicly provided by Trokielewicz et al. [24]. We have tested our models with all versions of Warsaw-BioBase-PostMortem-Iris database.

The Warsaw-BioBase-PostMortem-Iris-v1 dataset was gathered from subjects admitted to the hospital mortuary at the Medical University of Warsaw. Each eye was photographed using two distinct sensors: a professional handheld near-infrared (NIR) iris recognition camera (IriShield M2120U) and a consumer color camera (Olympus TG-3). The mortuary's temperature was maintained at approximately 42.8°F. Depending on tissue availability, images were captured over 2 to 8 acquisition sessions. Single-session images were obtained separately as per ISO/IEC 19795-2 guidelines, with the camera repositioned for each acquisition. The initial session consistently occurred 5-7 hours post-mortem, while subsequent sessions were less frequent and varied among subjects, as documented in the accompanying metadata. This dataset comprises 1330 post-mortem iris images from 17 individuals, captured at varying intervals after death, ranging from 5 hours to 17 days. This dataset includes both typical near-infrared (NIR) and high-quality visible light images, which we utilized for training our network.

1. Afzal Hossain; PhD Student, Electrical and Computer Engineering, Clarkson University, 8 Clarkson Avenue, Potsdam, NY13676, US, afhossa@clarkson.edu
2. Tipu Sultan; PhD Student, Mechanical and Aerospace Engineering, Clarkson University, 8 Clarkson Avenue, Potsdam, NY13676, US, sultant@clarkson.edu
3. Stephanie Schuckers; Professor, Electrical and Computer Engineering, Clarkson University, 8 Clarkson Avenue, Potsdam, NY13676, US, sschucke@clarkson.edu4

Notably, the post-mortem nature of the data is distinguishable by pronounced changes over time, including increased specular reflections due to tissue decay, corneal wrinkles, haze, altered pupil shape, and visible degradation of iris tissue along with partial collapses of the eyeball. For each sample in the dataset, the authors of [24] meticulously annotated a corresponding ground truth binary mask. This mask indicates regions of the iris unaffected by both post-mortem alterations and specular reflections, irrespective of their source. Figure 2 displays example images from the dataset alongside their respective binary ground truth masks.

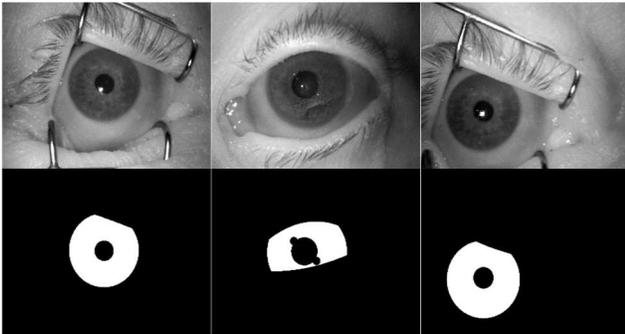

**Figure 2.** Example of post-mortem images from the Warsaw-BioBase-PostMortem-Iris-v1 dataset, along with their corresponding manually annotated masks.

### 4.2 Training and Evaluation Procedure:

For training and testing procedures, we use separate data which is already described in the above dataset section. We train the model with train portion of a dataset (14 out of 17) and then we test the model with test data (3 out of 17) and calculate the accuracy for each dataset. For the training, stochastic gradient descent is used as the minimization technique over 100 epochs in each experiment. We use default momentum, 0.001 learning rate, and 0.0005 L2 regularization. We apply data augmentation by Rescaling from range 0.9 to 1.1, Random Y-axis Reflection, Random X-axis Reflection, Random X-axis Translation and Random Y-axis Translation from range -15 to 15, and Random Rotation from range -20 to 20. Between the predicted masks and ground truth masks we calculate mean Intersection over Union (MeanIoU), mean boundary F1 score (F1 Score), ratio of the correctly classified pixels to the total number of pixels (Global Accuracy), mean percentage of correctly identified pixels for each class (Mean Accuracy), and average IoU of each class, weighted by the number of pixels in that class (Weighted IoU).

### 4.3 Results and Discussions

This study offers a comprehensive examination of iris segmentation techniques leveraging DL, with a particular emphasis on optimal architectural configurations and loss functions to improve segmentation accuracy. Our methodology involves employing two prominent semantic segmentation frameworks: SegNet and DeepLabV3+. We systematically explore various pre-trained neural network backbones, including VGG19, ResNet-18, ResNet-50, MobileNetV2, Xception, and InceptionResNetV2 ($M_1$ to $M_6$), to showcase their respective strengths. Notably, we introduce a significant modification by integrating the Dice loss function instead of the original loss functions of the DL models. Through rigorous experimentation on the Warsaw-BioBase-PostMortem-Iris-v1 dataset, we train all six DL models ($M_1$ to $M_6$) and calculate the segmentation accuracy mainly focusing on MeanIoU. Despite the implementation of the Dice loss function, the experimental analysis does not reveal significant improvements in performance metrics though Dice loss function and on MeanIoU are closely related and often used interchangeably. These findings carry significant implications for post-mortem iris segmentation techniques, offering valuable insights to inform and advance future endeavors in post-mortem iris segmentation and recognition systems.

In our investigation, we tried to find out the optimal image size for achieving superior accuracy in deep neural networks. We conducted experiments using images of dimensions 244x244, 256x256, and 640x480. Subsequently, we trained our ResNet-18 based DeepLabv3+ model to evaluate the accuracy corresponding to each image size. The experimental results, shown in Table 1, provide insights into the performance variations across different image dimensions. The table shows that in three instances, the 256x256 dimensional image provided the highest accuracy, while in two cases, the 224x224 dimensional image performed the best. Interestingly, the 640x480 dimensional image did not result in the highest accuracy in any accuracy metric. However, it is evident that for all accuracy metrics except the F1 score, the differences in accuracy are minimal, with variances of only one one-thousandth.

| Input Image Size | Global Accuracy (%) | Mean Accuracy (%) | Mean IoU (%) | Weighted IoU (%) | F1 Score (%) |
|---|---|---|---|---|---|
| 244x244 | 98.34 | **96.84** | 93.71 | 96.80 | **91.74** |
| 256x256 | **98.35** | 96.81 | **93.71** | **96.80** | 90.49 |
| 640x480 | 98.23 | 96.57 | 93.35 | 96.58 | 86.71 |

**Table 1.** Performance evaluation of ResNet-18 based DeepLabv3+ model across different image dimensions for Warsaw-BioBase-PostMortem-Iris-v1 dataset.

Observing Table 2, it's evident that our model $M_4$, comprising DeepLabv3+ with MobileNetV2, achieves the highest accuracy across all metrics except Mean Accuracy. Conversely, for Mean Accuracy, our model $M_6$, combining Inception-ResNet-v2 and DeepLabv3+, demonstrates the

1. Afzal Hossain; PhD Student, Electrical and Computer Engineering, Clarkson University, 8 Clarkson Avenue, Potsdam, NY13676, US, afhossa@clarkson.edu
2. Tipu Sultan; PhD Student, Mechanical and Aerospace Engineering, Clarkson University, 8 Clarkson Avenue, Potsdam, NY13676, US, sultant@clarkson.edu
3. Stephanie Schuckers; Professor, Electrical and Computer Engineering, Clarkson University, 8 Clarkson Avenue, Potsdam, NY13676, US, sschucke@clarkson.edu5

best performance. Our highest accuracy results for all accuracy metrices obtained from implementations utilizing the DeepLabv3+ semantic segmentation method, while the SegNet method did not yield the best accuracy in accuracy metric. In comparison to previous post-mortem iris segmentation method proposed by Trokielewicz et al. [24], which utilized a VGG16-based SegNet model, our proposed method $M_6$ (MobileNetV2 with DeepLabv3+) achieves superior segmentation performance. Trokielewicz et al. conducted training and testing on the Warsaw-BioBase-PostMortem-Iris-v1 dataset using a similar approach, employing 14 out of 17 subjects for training and the remaining 3 for testing. Their method yielded a MeanIoU of 88.53%, outperforming the conventional OSIRIS method, which achieved 73.58% MeanIoU. However, our proposed method $M_6$ achieved a significantly higher MeanIoU of 94.53%, outperforming OSIRIS by 20.95% and surpassing the method proposed by Trokielewicz et al. [24] by 6.00%.

| Method | Global Accuracy (%) | Mean Accuracy (%) | Mean IoU (%) | Weighted IoU (%) | F1 Score (%) |
|---|---|---|---|---|---|
| VGG19-SegNet ($M_1$) | 97.36 | 95.49 | 90.34 | 95.01 | 85.67 |
| ResNet18-DeepLabv3+ ($M_2$) | 98.35 | 96.81 | 93.71 | 96.80 | 90.49 |
| ResNet50-DeepLabv3+ ($M_3$) | 98.48 | 97.30 | 94.22 | 97.05 | 91.48 |
| MobileNetV2-DeepLabv3+ ($M_4$) | **98.57** | 97.22 | **94.53** | **97.22** | **91.98** |
| Xception-DeepLabv3+ ($M_5$) | 98.22 | 96.92 | 93.36 | 96.57 | 87.58 |
| InceptionResNetv2-DeepLabv3+ ($M_6$) | 98.34 | **97.53** | 93.84 | 96.81 | 88.16 |

**Table 2.** Performance evaluation of models $M_1$ to $M_6$ against Warsaw-BioBase-PostMortem-Iris-v1 dataset.

We have performed further experiments changing the last layer of the DL models. The last layer, common across all models, employs the cross-entropy loss function during training to optimize parameters. VGG19, ResNet18, ResNet50, MobileNetV2, Xception, and InceptionResNetV2 represent distinct architectures, each designed to balance performance and efficiency in different contexts. Notably, the utilization of the Pixel Classification Layer enables fine-grained classification at the pixel level, crucial for tasks such as semantic segmentation or object localization within images. We have replaced the last layer of the models with Dice loss function for as our primary emphasis in assessing accuracy metrics lies in MeanIoU, widely regarded as the most suitable metric for iris segmentation. Consequently, we opted to replace the loss function layer of pretrained DL models with the Dice loss function, given its close alignment with MeanIoU and interchangeable usage in this context. Table 3 illustrates the outcomes obtained from models employing the Dice loss function. Our model $M_4$, integrating DeepLabv3+ with MobileNetV2, emerges as the top performer across all metrics except Mean Accuracy. Conversely, for Mean Accuracy, our model $M_6$, leveraging the combination of Inception-ResNet-v2 and DeepLabv3+, showcases the highest performance. Comparing Table 2 and Table 3 reveals an interesting discovery that the models achieving the highest accuracy for different metrics are consistent. While the method achieving the overall best accuracy remains the same in both cases, replacing the loss function layer with the Dice loss function does not lead to improved accuracy for all methods. It appears that on average the pre-trained models perform better MeanIoU with their original loss functions instead of using the Dice loss function though the Dice loss function and the MeanIoU are closely related and often used interchangeably in image segmentation tasks due to their similarities in measuring the overlap between predicted and ground truth segmentations.

| Method | Global Accuracy (%) | Mean Accuracy (%) | Mean IoU (%) | Weighted IoU (%) | F1 Score (%) |
|---|---|---|---|---|---|
| VGG19-SegNet ($M_1$) | 97.32 | 95.80 | 90.63 | 95.76 | 85.52 |
| ResNet18-DeepLabv3+ ($M_2$) | 98.32 | 96.81 | 93.63 | 96.76 | 90.52 |
| ResNet50-DeepLabv3+ ($M_3$) | 98.44 | 96.52 | 94.01 | 96.97 | 90.44 |
| MobileNetV2-DeepLabv3+ ($M_4$) | **98.54** | 97.02 | **94.38** | **97.15** | **91.81** |
| Xception-DeepLabv3+ ($M_5$) | 98.20 | 96.76 | 93.27 | 96.53 | 87.47 |
| InceptionResNetv2-DeepLabv3+ ($M_6$) | 98.44 | **97.05** | 94.10 | 96.97 | 89.42 |

**Table 3.** Performance evaluation of models $M_1$ to $M_6$ replacing the last layer with Dice loss function against Warsaw-BioBase-PostMortem-Iris-v1 dataset.

1. Afzal Hossain; PhD Student, Electrical and Computer Engineering, Clarkson University, 8 Clarkson Avenue, Potsdam, NY13676, US, afhossa@clarkson.edu
2. Tipu Sultan; PhD Student, Mechanical and Aerospace Engineering, Clarkson University, 8 Clarkson Avenue, Potsdam, NY13676, US, sultant@clarkson.edu
3. Stephanie Schuckers; Professor, Electrical and Computer Engineering, Clarkson University, 8 Clarkson Avenue, Potsdam, NY13676, US, sschucke@clarkson.edu6

After discovering that the Dice loss function was not increasing the accuracy across all models, we decided to explore alternative loss functions by replacing the final layer of our top-performing model $M_4$. This model integrates DeepLabv3+ with MobileNetV2, utilizing cross-entropy loss function in its last classification layer. We replaced this layer with various loss functions and assessed their impact on key metrics such as Global Accuracy, Mean Accuracy, MeanIoU, Weighted IoU, and F1 Score. Our primary focus was on improving MeanIoU. Table 4 summarizes the overall accuracies achieved with different loss functions. Initially, we experimented with combining the Dice loss and cross-entropy loss, resulting in a MeanIoU of 94.42%. However, this combination did not yield significant accuracy improvements. Subsequently, we evaluated other loss functions to find out how much those can improve the accuracy. Employing the Lovász-Softmax loss function, we achieved a MeanIoU accuracy of 94.70%. By replacing the last layer of our model with the boundary loss function, we achieved a notable increase in MeanIoU, reaching 95.13%. Encouraged by this result, we explored a hybrid approach, combining the Boundary and Dice loss functions. This hybrid loss function provided the highest accuracy, with a MeanIoU of 95.54%. Figure 3 illustrates the comparison of MeanIoU obtained using original loss function and replacing the layer with other loss functions.

| Loss Function | Global Accuracy (%) | Mean Accuracy (%) | Mean IoU (%) | Weighted IoU (%) | F1 Score (%) |
|---|---|---|---|---|---|
| Dice Cross Entropy | 99.26 | 97.67 | 94.42 | 97.65 | 91.56 |
| Lovász-Softmax | 99.35 | 97.54 | 94.70 | 97.80 | 91.74 |
| Boundary | 99.47 | 97.53 | 95.13 | 98.03 | 92.10 |
| Boundary Dice | **99.57** | **98.40** | **95.54** | **98.22** | **93.05** |

**Table 4:** Performance evaluation of model MobileNetV2 with DeepLabv3+ ($M_4$), replacing the last layer with different loss functions against Warsaw-BioBase-PostMortem-Iris-v1 dataset.

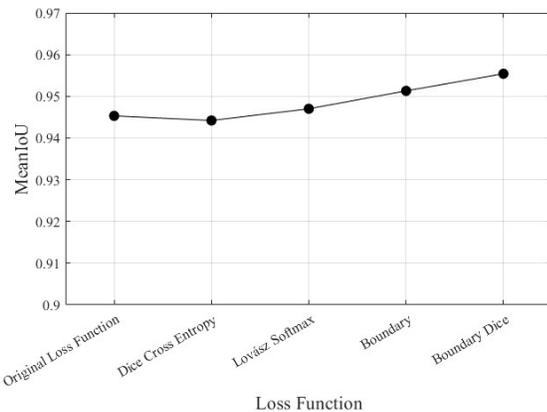

**Figure 3.** The MeanIoU values for the method MobileNetV2-DeepLabv3+ ($M_4$) with original loss function and replacing it with other loss functions for Warsaw-BioBase-PostMortem-Iris-v1 dataset.

We conducted cross-validation of our approach by evaluating its performance on the Warsaw-BioBase-PostMortem-Iris-v2 and Warsaw-BioBase-PostMortem-Iris-v3 datasets. To accomplish this, we utilized our iris segmentation toolkit to segment the iris images. Our methodology incorporates MobileNetv2 as the backbone of DeepLabV3+ and replaces the final layer with a hybrid loss function that combines Boundary and Dice loss. This method is already achieving a MeanIoU of 95.54% on the Warsaw-BioBase-PostMortem-Iris-v1 dataset. Our proposed method consistently achieves over 90% MeanIoU on additional datasets. Table 5 presents the experimental results.

| Dataset | Global Accuracy (%) | Mean Accuracy (%) | Mean IoU (%) | Weighted IoU (%) | F1 Score (%) |
|---|---|---|---|---|---|
| Warsaw-BioBase-PostMortem-Iris-v2 | 96.43 | 95.66 | 92.57 | 92.73 | 90.31 |
| Warsaw-BioBase-PostMortem-Iris-v3 | 94.58 | 93.35 | 90.39 | 90.31 | 88.03 |

**Table 5.** Cross validation check of our evaluation of our method against Warsaw-BioBase-PostMortem-Iris-v2 and Warsaw-BioBase-PostMortem-Iris-v3 datasets.

## 5. Conclusion

This paper explores the application of DL models for segmenting iris images from deceased individuals, a challenging task in forensic analysis. By training SegNet and DeepLabV3+ with various backbone architectures, including MobileNetv2, the study achieves significant improvements in accuracy. We explore optimizing architectural configurations and replacing the last layer of the DL models with a variety of loss functions to enhance accuracy. Using a hybrid loss function (combination of Boundary and Dice loss) and MobileNetv2 as a backbone of DeepLabV3+, our proposed method achieves a MeanIoU of 95.54% on the Warsaw-BioBase-PostMortem-Iris-v1 dataset. Cross-validation on additional datasets validates the method's effectiveness, consistently achieving over 90% MeanIoU. These findings advance post-mortem iris

1. Afzal Hossain; PhD Student, Electrical and Computer Engineering, Clarkson University, 8 Clarkson Avenue, Potsdam, NY13676, US, afhossa@clarkson.edu
2. Tipu Sultan; PhD Student, Mechanical and Aerospace Engineering, Clarkson University, 8 Clarkson Avenue, Potsdam, NY13676, US, sultant@clarkson.edu
3. Stephanie Schuckers; Professor, Electrical and Computer Engineering, Clarkson University, 8 Clarkson Avenue, Potsdam, NY13676, US, sschucke@clarkson.edu7

segmentation techniques, offering insights for future development. Our work representing the most extensive evaluation of DNNs for post-mortem iris segmentation to date.

## References

[1] C. Li, W. Zhou, and S. Yuan, "Iris recognition based on a novel variation of local binary pattern," *Vis Comput*, vol. 31, no. 10, pp. 1419–1429, Oct. 2015, doi: 10.1007/s00371-014-1023-5.

[2] J. Daugman, "Chapter 25 - How Iris Recognition Works," in *The Essential Guide to Image Processing*, A. Bovik, Ed., Boston: Academic Press, 2009, pp. 715–739. doi: 10.1016/B978-0-12-374457-9.00025-1.

[3] P. Commission and others, "Government of India, Unique Identification Authority of India." 2012.

[4] J. Daugman, "Iris Recognition at Airports and Border Crossings." 2009.

[5] A. Kuehlkamp et al., "Interpretable Deep Learning-Based Forensic Iris Segmentation and Recognition," presented at the Proceedings of the IEEE/CVF Winter Conference on Applications of Computer Vision, 2022, pp. 359–368. Accessed: Apr. 08, 2024. [Online]. Available: https://openaccess.thecvf.com/content/WACV2022W/XAI4B/html/Kuehlkamp_Interpretable_Deep_Learning-Based_Forensic_Iris_Segmentation_and_Recognition_WACVW_2022_paper.html

[6] P. J. Grother, J. R. Matey, E. Tabassi, G. W. Quinn, and M. Chumakov, "IREX VI - Temporal Stability of Iris Recognition Accuracy," *NIST*, Jul. 2013, Accessed: Apr. 05, 2024. [Online]. Available: https://www.nist.gov/publications/irex-vi-temporal-stability-iris-recognition-accuracy

[7] M. Trokielewicz, A. Czajka, and P. Maciejewicz, "Human iris recognition in post-mortem subjects: Study and database," in *2016 IEEE 8th International Conference on Biometrics Theory, Applications and Systems (BTAS)*, Sep. 2016, pp. 1–6. doi: 10.1109/BTAS.2016.7791175.

[8] "The Effect of Decomposition on the Efficacy of Biometrics for Positive Identification - Sauerwein - 2017 - Journal of Forensic Sciences - Wiley Online Library." Accessed: Apr. 05, 2024. [Online]. Available: https://onlinelibrary.wiley.com/doi/full/10.1111/1556-4029.13484

[9] A. Czajka and K. W. Bowyer, "Presentation Attack Detection for Iris Recognition: An Assessment of the State-of-the-Art," *ACM Comput. Surv.*, vol. 51, no. 4, p. 86:1-86:35, Jul. 2018, doi: 10.1145/3232849.

[10] M. Trokielewicz, A. Czajka, and P. Maciejewicz, "Post-mortem iris recognition with deep-learning-based image segmentation," *Image and Vision Computing*, vol. 94, p. 103866, Feb. 2020, doi: 10.1016/j.imavis.2019.103866.

[11] A. k H. Sansola, "Postmortem iris recognition and its application in human identification," M.S., Boston University, United States -- Massachusetts, 2015. Accessed: Apr. 05, 2024. [Online]. Available: https://www.proquest.com/docview/1732874333/abstract/BE2401929D4608PQ/1

[12] M. Trokielewicz, A. Czajka, and P. Maciejewicz, "Post-mortem human iris recognition," in *2016 International Conference on Biometrics (ICB)*, Jun. 2016, pp. 1–6. doi: 10.1109/ICB.2016.7550073.

[13] H. Hofbauer, F. Alonso-Fernandez, J. Bigun, and A. Uhl, "Experimental analysis regarding the influence of iris segmentation on the recognition rate," *IET Biometrics*, vol. 5, no. 3, pp. 200–211, 2016, doi: 10.1049/iet-bmt.2015.0069.

[14] A. Szczepański, K. Misztal, and K. Saeed, "Pupil and Iris Detection Algorithm for Near-Infrared Capture Devices," in *Computer Information Systems and Industrial Management*, K. Saeed and V. Snášel, Eds., Berlin, Heidelberg: Springer, 2014, pp. 141–150. doi: 10.1007/978-3-662-45237-0_15.

[15] A. Boyd et al., "Post-Mortem Iris Recognition—A Survey and Assessment of the State of the Art," *IEEE Access*, vol. 8, pp. 136570–136593, 2020, doi: 10.1109/ACCESS.2020.3011364.

[16] S. K. Saripalle, A. McLaughlin, R. Krishna, A. Ross, and R. Derakhshani, "Post-mortem iris biometric analysis in Sus scrofa domesticus," in *2015 IEEE 7th International Conference on Biometrics Theory, Applications and Systems (BTAS)*, Sep. 2015, pp. 1–5. doi: 10.1109/BTAS.2015.7358789.

[17] D. S. Bolme, R. A. Tokola, C. B. Boehnen, T. B. Saul, K. A. Sauerwein, and D. W. Steadman, "Impact of environmental factors on biometric matching during human decomposition," in *2016 IEEE 8th International Conference on Biometrics Theory, Applications and Systems (BTAS)*, Sep. 2016, pp. 1–8. doi: 10.1109/BTAS.2016.7791177.

[18] M. Trokielewicz, A. Czajka, and P. Maciejewicz, "Iris Recognition After Death," *IEEE Transactions on Information Forensics and Security*, vol. 14, no. 6, pp. 1501–1514, Jun. 2019, doi: 10.1109/TIFS.2018.2881671.

[19] A. Krizhevsky, I. Sutskever, and G. E. Hinton, "ImageNet Classification with Deep Convolutional Neural Networks," in *Advances in Neural Information Processing Systems*, Curran Associates, Inc., 2012. Accessed: Sep. 10, 2023. [Online]. Available: https://proceedings.neurips.cc/paper/2012/hash/c399862d3b9d6b76c8436e924a68c45b-Abstract.html

[20] A. Garcia-Garcia, S. Orts-Escolano, S. Oprea, V. Villena-Martinez, and J. Garcia-Rodriguez, "A Review on Deep Learning Techniques Applied to Semantic Segmentation." arXiv, Apr. 22, 2017. doi: 10.48550/arXiv.1704.06857.

[21] R. P. Broussard and R. W. Ives, "Using artificial neural networks and feature saliency to identify iris measurements that contain the most discriminatory information for iris segmentation," in *2009 IEEE Workshop on Computational Intelligence in Biometrics: Theory, Algorithms, and Applications*, Mar. 2009, pp. 46–51. doi: 10.1109/CIB.2009.4925685.

[22] F. Marra, G. Poggi, C. Sansone, and L. Verdoliva, "A deep learning approach for iris sensor model identification," *Pattern Recognition Letters*, vol. 113, pp. 46–53, Oct. 2018, doi: 10.1016/j.patrec.2017.04.010.

[23] N. Liu, H. Li, M. Zhang, J. Liu, Z. Sun, and T. Tan, "Accurate iris segmentation in non-cooperative environments using fully convolutional networks," in *2016 International Conference on Biometrics (ICB)*, Jun. 2016, pp. 1–8. doi: 10.1109/ICB.2016.7550055.

1. Afzal Hossain; PhD Student, Electrical and Computer Engineering, Clarkson University, 8 Clarkson Avenue, Potsdam, NY13676, US, afhossa@clarkson.edu
2. Tipu Sultan; PhD Student, Mechanical and Aerospace Engineering, Clarkson University, 8 Clarkson Avenue, Potsdam, NY13676, US, sultant@clarkson.edu
3. Stephanie Schuckers; Professor, Electrical and Computer Engineering, Clarkson University, 8 Clarkson Avenue, Potsdam, NY13676, US, sschucke@clarkson.edu8

[24] M. Trokielewicz and A. Czajka, "Data-driven segmentation of post-mortem iris images," in *2018 International Workshop on Biometrics and Forensics (IWBF)*, Jun. 2018, pp. 1–7. doi: 10.1109/IWBF.2018.8401558.

[25] V. Badrinarayanan, A. Kendall, and R. Cipolla, "SegNet: A Deep Convolutional Encoder-Decoder Architecture for Image Segmentation," *IEEE Transactions on Pattern Analysis and Machine Intelligence*, vol. 39, no. 12, pp. 2481–2495, Dec. 2017, doi: 10.1109/TPAMI.2016.2644615.

[26] L.-C. Chen, Y. Zhu, G. Papandreou, F. Schroff, and H. Adam, "Encoder-Decoder with Atrous Separable Convolution for Semantic Image Segmentation," presented at the Proceedings of the European Conference on Computer Vision (ECCV), 2018, pp. 801–818. Accessed: Sep. 10, 2023. [Online]. Available: https://openaccess.thecvf.com/content_ECCV_2018/html/Liang-Chieh_Chen_Encoder-Decoder_with_Atrous_ECCV_2018_paper.html

[27] L.-C. Chen, G. Papandreou, F. Schroff, and H. Adam, "Rethinking Atrous Convolution for Semantic Image Segmentation." arXiv, Dec. 05, 2017. doi: 10.48550/arXiv.1706.05587.

[28] L.-C. Chen, G. Papandreou, I. Kokkinos, K. Murphy, and A. L. Yuille, "DeepLab: Semantic Image Segmentation with Deep Convolutional Nets, Atrous Convolution, and Fully Connected CRFs," *IEEE Transactions on Pattern Analysis and Machine Intelligence*, vol. 40, no. 4, pp. 834–848, Apr. 2018, doi: 10.1109/TPAMI.2017.2699184.

[29] K. Simonyan and A. Zisserman, "Very Deep Convolutional Networks for Large-Scale Image Recognition." arXiv, Apr. 10, 2015. doi: 10.48550/arXiv.1409.1556.

[30] K. He, X. Zhang, S. Ren, and J. Sun, "Deep Residual Learning for Image Recognition," presented at the Proceedings of the IEEE Conference on Computer Vision and Pattern Recognition, 2016, pp. 770–778. Accessed: Sep. 10, 2023. [Online]. Available: https://openaccess.thecvf.com/content_cvpr_2016/html/He_Deep_Residual_Learning_CVPR_2016_paper.html

[31] M. Sandler, A. Howard, M. Zhu, A. Zhmoginov, and L.-C. Chen, "MobileNetV2: Inverted Residuals and Linear Bottlenecks," presented at the Proceedings of the IEEE Conference on Computer Vision and Pattern Recognition, 2018, pp. 4510–4520. Accessed: Sep. 10, 2023. [Online]. Available: https://openaccess.thecvf.com/content_cvpr_2018/html/Sandler_MobileNetV2_Inverted_Residuals_CVPR_2018_paper.html

[32] F. Chollet, "Xception: Deep Learning With Depthwise Separable Convolutions," presented at the Proceedings of the IEEE Conference on Computer Vision and Pattern Recognition, 2017, pp. 1251–1258. Accessed: Sep. 10, 2023. [Online]. Available: https://openaccess.thecvf.com/content_cvpr_2017/html/Chollet_Xception_Deep_Learning_CVPR_2017_paper.html

[33] C. Szegedy, S. Ioffe, V. Vanhoucke, and A. Alemi, "Inception-v4, Inception-ResNet and the Impact of Residual Connections on Learning," *Proceedings of the AAAI Conference on Artificial Intelligence*, vol. 31, no. 1, Art. no. 1, Feb. 2017, doi: 10.1609/aaai.v31i1.11231.

[34] "Generalized Overlap Measures for Evaluation and Validation in Medical Image Analysis | IEEE Journals & Magazine | IEEE Xplore." Accessed: Apr. 05, 2024. [Online]. Available: https://ieeexplore.ieee.org/abstract/document/1717643

[35] C. H. Sudre, W. Li, T. Vercauteren, S. Ourselin, and M. Jorge Cardoso, "Generalised Dice Overlap as a Deep Learning Loss Function for Highly Unbalanced Segmentations," in *Deep Learning in Medical Image Analysis and Multimodal Learning for Clinical Decision Support*, M. J. Cardoso, T. Arbel, G. Carneiro, T. Syeda-Mahmood, J. M. R. S. Tavares, M. Moradi, A. Bradley, H. Greenspan, J. P. Papa, A. Madabhushi, J. C. Nascimento, J. S. Cardoso, V. Belagiannis, and Z. Lu, Eds., Cham: Springer International Publishing, 2017, pp. 240–248. doi: 10.1007/978-3-319-67558-9_28.

[36] F. Milletari, N. Navab, and S.-A. Ahmadi, "V-Net: Fully Convolutional Neural Networks for Volumetric Medical Image Segmentation," in *2016 Fourth International Conference on 3D Vision (3DV)*, Oct. 2016, pp. 565–571. doi: 10.1109/3DV.2016.79.

1. Afzal Hossain; PhD Student, Electrical and Computer Engineering, Clarkson University, 8 Clarkson Avenue, Potsdam, NY13676, US, afhossa@clarkson.edu
2. Tipu Sultan; PhD Student, Mechanical and Aerospace Engineering, Clarkson University, 8 Clarkson Avenue, Potsdam, NY13676, US, sultant@clarkson.edu
3. Stephanie Schuckers; Professor, Electrical and Computer Engineering, Clarkson University, 8 Clarkson Avenue, Potsdam, NY13676, US, sschucke@clarkson.edu9